\documentclass[conference]{IEEEtran}
\usepackage{amsmath,amsfonts}
\usepackage{algorithm}
\usepackage{adjustbox}
\usepackage{lscape}
\usepackage{siunitx}
\usepackage[noend]{algpseudocode}
\usepackage{textcomp}
\usepackage{soul}
\usepackage{booktabs}
\usepackage[utf8]{inputenc}
\usepackage[T1]{fontenc}
\usepackage{rotating}
\usepackage{graphicx}
\usepackage{paralist}
\usepackage{tabularx}
\usepackage{multicol}
\usepackage{multirow}
\usepackage{pbox}
\usepackage{enumitem}	
\usepackage{colortbl}
\usepackage{pifont}
\usepackage{xspace}
\usepackage{url}
\usepackage{tikz}
\usepackage{tabularx}
\usepackage{fontawesome}
\usepackage{lscape}
\usepackage{listings}
\usepackage{color}
\usepackage{anyfontsize}
\usepackage{comment}
\usepackage{soul}
\usepackage{multibib}
\usepackage{multirow}
\usepackage{xspace}
\usepackage{caption}
\usepackage{footnote}
\usepackage{tcolorbox}
\usepackage{booktabs}
\usepackage{fixltx2e}
\usepackage{balance}
\usepackage[normalem]{ulem}
\usepackage{graphicx}

\definecolor{Gray}{gray}{0.9}
\definecolor{codegreen}{rgb}{0,0.6,0}
\definecolor{codegray}{rgb}{0.73,0.38,0.06}
\definecolor{codepurple}{rgb}{0.70,0.27,0}
\definecolor{codemagenta}{rgb}{0.74,0.09,0.42}
\definecolor{codeoutput}{rgb}{0.5,0,0}
\definecolor{backcolour}{rgb}{0.96,0.96,0.96}

\newboolean{showcomments}

\setboolean{showcomments}{true}

\ifthenelse{\boolean{showcomments}}
{\newcommand{\nb}[2]{
		\fbox{\bfseries\sffamily\scriptsize#1}
		{\sf\small$\blacktriangleright$\textit{#2}$\blacktriangleleft$}
	}
	
}
{\newcommand{\nb}[2]{}
	
}

\newcommand{\ie}{\emph{i.e.,}\xspace}
\newcommand{\eg}{\emph{e.g.,}\xspace}
\newcommand{\etc}{etc.\xspace}
\newcommand{\etal}{\emph{et~al.}\xspace}
\newcommand{\secref}[1]{Section~\ref{#1}\xspace}
\newcommand{\figref}[1]{Fig.~\ref{#1}\xspace}

\newcommand{\tabref}[1]{Table~\ref{#1}\xspace}

\newcommand{\contexts}{8\xspace}

\begin{document}

\title{Deep Learning-based Code Completion: On the Impact on Performance of Contextual Information}

\author{
\IEEEauthorblockN{Matteo Ciniselli\IEEEauthorrefmark{1}, Luca Pascarella\IEEEauthorrefmark{2}, Gabriele Bavota\IEEEauthorrefmark{1}}
\IEEEauthorblockA{\IEEEauthorrefmark{1}\textit{SEART @ Software Institute, Universit\`{a} della Svizzera italiana (USI), Switzerland}}
\IEEEauthorblockA{\IEEEauthorrefmark{2}\textit{PBL Group @ ETH Zurich, Switzerland}}
}


\maketitle

\begin{abstract}
Code completion aims at speeding up code writing by recommending to developers the next tokens they are likely to type. Deep Learning (DL) models pushed the boundaries of code completion by \emph{redefining} what these coding assistants can do: We moved from predicting few code tokens to automatically generating entire functions. One important factor impacting the performance of DL-based code completion techniques is the \emph{context} provided them as input. With ``context'' we refer to \emph{what the model knows} about the code to complete. In a simple scenario, the DL model might be fed with a partially implemented function to complete. In this case, the context is represented by the incomplete function and, based on it, the model must generate a prediction. It is however possible to expand such a context to include additional information, like the whole source code file containing the function to complete, which could be useful to boost the prediction performance. In this work, we present an empirical study investigating how the performance of a DL-based code completion technique is affected by different contexts. We experiment with \contexts types of contexts and their combinations. These contexts include: (i) \emph{coding contexts}, featuring information extracted from the code base in which the code completion is invoked (\eg code components structurally related to the one to ``complete''); (ii) \emph{process context}, with information aimed at depicting the current status of the project in which a code completion task is triggered (\eg a textual representation of open issues relevant for the code to complete); and (iii) \emph{developer contexts}, capturing information about the developer invoking the code completion (\eg the APIs they frequently use). Our results show that additional contextual information can benefit the performance of DL-based code completion, with relative improvements up to +22\% in terms of correct predictions.
\end{abstract}

\begin{IEEEkeywords}
Code Completion, DL4SE, Empirical Study
\end{IEEEkeywords}

\section{Introduction} \label{sec:intro}

One of the most noticeable results of the adoption of Deep Learning (DL) in software engineering (SE) is the recent release of DL-based programming assistants such as GitHub Copilot \cite{copilot}. These tools \emph{redefined} the notion of code completion, moving it from techniques able to recommend the next few tokens the developer is likely to type to tools capable of automatically generating entire functions. While Copilot likely is the most well-known representative of this generation of code completion tools, it is the natural follow-up of years of research done in this field \cite{Perelman:pldi2012,GveroKKP13,kyaw2018proposal,wang2020towards,ciniselli:tse2021}. Most of these works proposed novel solutions with the main goal of improving the state-of-the-art performance of code completion tools. When talking about ``performance'', we refer to the accuracy of the technique in recommending the expected code.

When it comes to DL-based code completion tools, one important factor affecting their performance is the contextual information provided as input to the model for triggering the recommendation: This is the information available to the model to decide which code completion recommendation to generate. For example, in the recent work by Ciniselli \etal \cite{ciniselli:tse2021} the DL model is provided as input a Java method with one or more missing statements to complete. 

In this case, the incomplete Java method is the only information the model can rely upon to predict the missing statements. Such a design choice ensures shorter inputs for the model and, as a consequence, shorter training time. However, this choice may limit the prediction capability of the model which could benefit, for example, from knowing what the other methods implemented in the same class are. While previous work suggest the positive impact that additional contextual information may have when using DL-based solutions for code-related tasks \cite{tian:icsme2022}, little is known about the role of contextual information on the prediction accuracy of DL-based code completion techniques. This is the focus of our work.

We start by defining three families of contextual information which can be provided to a DL model to improve its prediction capabilities. To provide a high-level explanation of each of them, let us focus on the same method-level code completion task defined by Ciniselli \etal and previously summarized (\ie provide as input an incomplete Java method and ask the model to generate the missing part). We use $IM_i$ to refer to a generic incomplete method to finalize and assume that the developer $D_j$ is the one working on it (\ie the person who will receive the completion recommendation). The first family of contextual information we experiment with is the \emph{coding context}: These are contexts augmenting the input provided to the model with code components having structural relationships with $IM_i$ (\eg the methods invoking/invoked by $IM_i$). The assumption is that knowing more about the code base can help the model in generating the correct prediction.  

The second family is the \emph{process context}, providing the model with information related to the development process carried out in the project $IM_i$ belongs to (\eg what the open issues possibly related to $IM_i$ are). The idea behind this context is that knowing the ongoing tasks could help the model in predicting the missing code. 
Finally, the third family is the \emph{developer context}, augmenting the input with information characterizing the recent development activity of $D_j$ (\ie the developer currently working on $IM_i$). 

One of the developer contexts we experiment with augments the model's input with method invocations recently and frequently used by $D_j$. The assumption is that the model might exploit information about the recent development activities of $D_j$ to improve its predictions. 

We defined \contexts types of contexts belonging to the three above-described families and experimented them (and their combinations) by training and testing 18 Text-To-Text-Transfer-Transformer (T5) \cite{t5} models. We show that additional contextual information helps in boosting prediction performance, with the ones belonging to the \emph{coding context} bringing the larger boost. By combining different types of contexts it is possible to achieve a relative improvement of up to +22\%.

\section{Types of Contextual Information} \label{sec:contexts}
This section introduces the types of contextual information we experiment with. They are all depicted in \figref{fig:context} which does also include what we refer to as ``baseline'' (see red part in \figref{fig:context}). The baseline represents the common code completion scenario experimented in the literature, in which the DL model is only fed with the piece of code to complete. We adopt the method-level completion recently experimented by Ciniselli \etal \cite{ciniselli:tse2021} in which one or more statements are masked in a Java method (see the \texttt{$<$MISSING CODE$>$} tag \figref{fig:context}) with the model in charge of predicting them. The baseline will be used to assess the boost in performance (if any) provided by the additional contextual information provided to the model. The examples in \figref{fig:context} are extracted from a real instance present in the training dataset that will be described in \secref{sec:dataset}. 

In the following we use $IM_i$ to refer to the incomplete method provided to the model (\ie \texttt{handle\-Data\-Process\-Exception} in \figref{fig:context}). All contexts we describe represent additional information that is provided to the model on top of the ``baseline'' representation. 

The goal of this section is not to provide all technical details about \emph{how} we create these contexts, but rather to present and justify them. Technicalities about how we built the different datasets needed to experiment with these contexts are presented in \secref{sec:dataset}.

We experiment with three families of contexts: \emph{coding} (green in \figref{fig:context}), \emph{process} (blue), and \emph{developer} (yellow).

\subsection{Coding Context} \label{sec:context:coding}
The basic idea is to augment what the model knows about the code to complete with additional information extracted from the code base. We devise three types of coding contexts. 

The first, \textbf{method calls}, provides the model with the complete signature of the methods invoked by or invoking the method to complete ($IM_i$). Note that, being $IM_i$ an ``artificial'' incomplete method we created by replacing some of its statements with the \texttt{$<$MISSING CODE$>$} tag, methods that were invoked in the replaced statements are not included in the context. Indeed, in a real usage scenario those statements would not exist. In \figref{fig:context} it can be seen that we use the special tag \texttt{$<$OUT$>$} to mark methods invoked by $IM_i$ and \texttt{$<$IN$>$} for methods invoking $IM_i$. 

The rationale behind this context is that the model might benefit from knowing the code components structurally coupled (via method calls) to $IM_i$. It is important to discuss at this point why we only include the signatures of the coupled methods rather than their full implementation which, in theory, could provide even more information to the model. Such a choice is due to technical limitations of the DL models, which are able to deal with input sequences of limited length. 

For example, recent work in the SE literature capped the input instances to 512 tokens \cite{guo:iclr2021,yang:seke2019,yang:seke2019a,ciniselli:msr2021,alon:iclr2019,ciniselli:tse2021,aye:icse2021}. We pushed this boundary to 1,024 tokens which still limits the contextual information size. Thus, in all contexts, we consider such a tradeoff between the amount of additional information and the size of the input sequence.

The second coding context we define is the \textbf{class signatures}, providing the model with the signature of all other methods contained in the class implementing $IM_i$ (bottom-left corner of \figref{fig:context}). The rationale behind this context is that, accordingly to the ``high cohesion'' principle \cite{Constantine:ibm1974}, classes are supposed to group together related methods. 

Finally, \textbf{most similar method} is the third coding context we defined. When experimenting with DL models for code completion there must be no duplicates between the training and the test datasets. Otherwise, the model would be asked to complete methods it has already seen during training, thus artificially inflating its performance. However, it is reasonable to think that, given an incomplete method  $IM_i$ from the test set, a ``similar'' method may exist in the training set. A concrete example from our dataset is depicted in the bottom-right corner of \figref{fig:context}: The method \texttt{connection\-Close} from our training set is the most similar to \texttt{handle\-Data\-Process\-Exception} (our $IM_i$). Indeed, it can be seen that they share some logic. Our assumption is that such an additional contextual information can help the model in predicting the missing statements. Note that we retrieve the most similar method \emph{only from the training set}. This is important since, in a real scenario, the instances in the training set are the only ones known to the model. We detail in \secref{sec:dataset} how the most similar method is identified. 

\subsection{Process Context} \label{sec:context:process}

The process contexts provide the model with information capturing ``what is going on'' in the project when a recommendation must be triggered to complete $IM_i$. We define two types of process contexts, both exploiting information from the issue tracker of the project $IM_i$ belongs to. The assumption is that if a developer is implementing code aimed at addressing an open issue, information extracted from such an issue may help the model in recommending the needed code. An issue is usually composed by a \emph{title}, which summarizes its content, and a \emph{body}, which provides a more detailed description of the issue.
These two elements are the ones driving the definition of our two contexts, named \textbf{issue title} and \textbf{issue body} (see \figref{fig:context}). For both of them we start by identifying for the given $IM_i$ the most similar open issue at the time $t$ the code completion on $IM_i$ is invoked. 

\begin{landscape}\centering
\vspace*{\fill}
\begin{figure}[ht]
   \centering
    \includegraphics[width=1.2\textwidth]{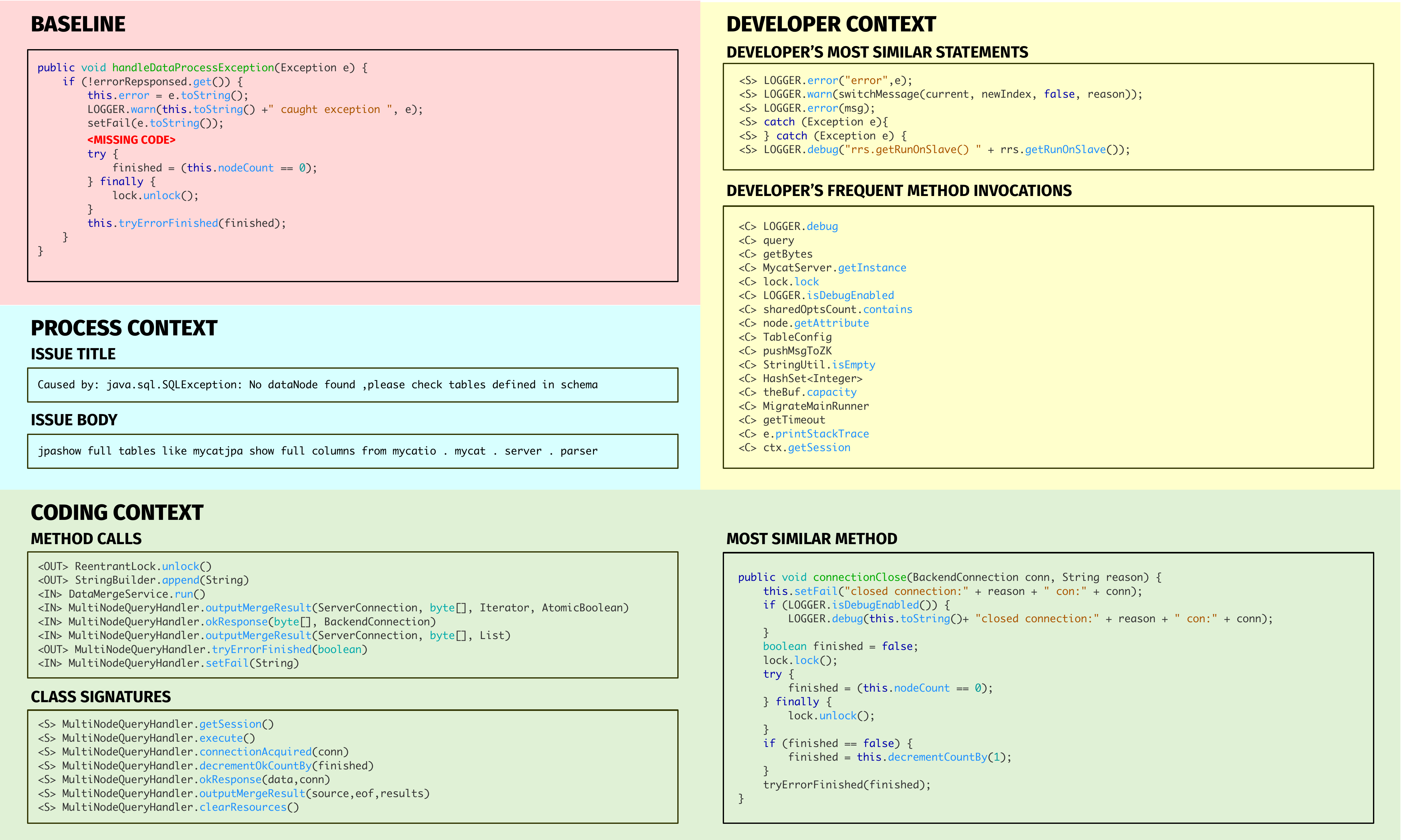}
    \caption{Experimented contexts: Examples from a real instance in our dataset.}
    \label{fig:context}
\end{figure}
\vfill
\end{landscape}

We use such an open issue to create the two contexts, one featuring the issue title and the other one the issue body. While a context featuring their combination would make sense, this is not presented since in our experiment we test with several combinations of contexts both from the same family and from different families (\ie a process context mixed up with a coding context). 

\subsection{Developer Context}  \label{sec:context:developer}

The last family of contexts provides the DL model with information about the developer $D_j$ currently working on $IM_i$ (\ie the one receiving the completion recommendation). The idea is that different developers may have different coding styles \cite{corbo2007smart,hokka2021linking} (\eg may favor the usage of specific APIs).

The first context in this family is the \textbf{developer's most similar statements} (top-right corner in \figref{fig:context}). We mine the recent commits performed by $D_j$ and extract from each of them the source code statements they added, deleted, or edited. Each statement in this set is then compared with the method to complete $IM_i$ to identify the statements being more similar to the code under development. The most similar statements are added as contextual information to the model input, separated by an \texttt{$<$S$>$} tag as shown in \figref{fig:context}.


The second \emph{developer context} is the \textbf{developer's frequent meth\-od invocations}, providing the model with information about meth\-od calls (both internal and external to the project) recently and frequently used by $D_j$, as depicted in \figref{fig:context}. Technical details about the building of the two developer contexts are presented in \secref{sec:dataset}.

\section{Building the ``context datasets''} \label{sec:dataset}

Experimenting with the contexts described in \secref{sec:contexts} requires the building of several datasets aimed at training/testing the DL model to assess the impact of the different contextual information on its performance. The first step in this process is the selection of software repositories from which the information needed for the different contexts can be extracted. We used the tool by Dabic \etal~\cite{Dabic:msr2021data} to select all GitHub non-forked Java repositories having more than 100 commits, 10 contributors, 50 issues, and 10 stars. These filters ensure that the selected projects (i) have a substantial history, needed to extract information related to the \emph{developer context}; (ii) are not personal/toy projects, since at least 10 contributors took part to their development; and (iii) use the issue tracker, a requirement for the extraction of the \emph{process context}. The 10-star filter is mandatory in the search tool by Dabic \etal~\cite{Dabic:msr2021data}, since projects with less than 10 stars are not indexed. The result of this query was a list of 5,632 candidate repositories. The extraction of the coding context (\eg the identification of the methods invoked by or invoking the method of interest) requires the source code of the selected projects to be compilable. For this reason, we excluded projects do not explicitly tagging releases in GitHub and, for the remaining ones, we checked-out their last two releases trying to compile them using Maven or Gradle. 

\eject

The choice of focusing on the last two releases rather than only on the last one aimed at increasing the number of projects suitable for our study. Through this process, we successfully compiled at least one release for 1,072 repositories. For each of them, we indicate with $R_c$ the successfully compiled release.

We use these repositories to build eight training/testing datasets, one for the baseline approach emulating the work by Ciniselli \etal~\cite{ciniselli:tse2021} and one for each of the seven contexts introduced in \secref{sec:contexts}. It is important to remember that all datasets will feature instances in the form $IM_i \rightarrow C_r$, where $IM_i$ represents an incomplete method (\ie a method from which specific statements have been masked) and $C_r$ represents the expected completion (\ie the code the model should generate). $C_r$ is identical across all eight datasets, while $IM_i$ changes based on the contextual representation it features. 

We start by checking-out all 1,072 compilable releases from which we randomly selected up to 1,000 Java files per project. The cap at 1,000 files per repository has been defined to avoid very large repositories to influence too much the final dataset (\eg to contribute 50\% of the final instances). Also, when selecting those files, we ignored those containing the word ``test'' in their name or belonging to a package having ``test'' in their path. This was done in an attempt to exclude test files, thus building a more cohesive dataset only featuring production code instances. From each of the selected files, we extracted the implemented methods using \textsc{javalang} \cite{javalang}. We then removed all methods exceeding 682 tokens. While this may look like a \emph{magic number}, it represents two-thirds of the space available for the model's input. Indeed, as detailed in \secref{sec:design}, our DL model accepts inputs up to 1,024 tokens in length. We decided to dedicate up to 682 tokens to the representation of $IM_i$ and at least 342 tokens for representing the additional contextual information. The contextual information is appended to $IM_i$. Thus, in case the contextual information makes the input longer than 1,024 tokens, part of it will be cut and ignored by the model. For example, if $IM_i$ requires 550 tokens and a specific type of contextual information requires 750 tokens, the last 276 tokens of the context will not be seen by the model. 

The above-described process resulted in 42,182 collected methods. Ciniselli \etal \cite{ciniselli:tse2021} masked randomly selected statements in methods to create the instances $IM_i \rightarrow C_r$. We decided to adopt a different approach for the masking, with the goal of better simulating a developer writing code and receiving completion recommendations. In particular, rather than masking randomly selected statements, we run \texttt{git blame} on each method in our dataset, retrieving the latest commit before $R_c$ (\ie the compiled release) which changed at least one code statement in the method. Let us assume that the identified commit changes a single statement $s_k$: We create an instance $IM_i \rightarrow C_r$ in which $IM_i$ has $s_k$ masked and $C_r = s_k$. Such an instance simulates a realistic change which has been actually performed in the history of the project. It could also happen that the identified commit changes several statements in the method. 

\eject

To limit the complexity of the code completion problem, we decided to mask at most two complete statements when creating an $IM_i$. Thus, if five statements have been modified ($s_1$, $s_2$, $s_3$, $s_4$, and $s_5$), we create three $IM_i \rightarrow C_r$ instances from the corresponding method: The first has $s_1$ and $s_2$ masked, the second has $s_3$ and $s_4$ masked, the third has only $s_5$ masked. 

This process generates the ``baseline'' dataset, in which the $IM_i$ in the instances is only composed by the incomplete method, with no additional contextual information. In the following we describe how we built the remaining seven datasets. We ensure no duplicates in our datasets, removing instances having identical $IM_i$. In order to fairly compare the performance of the DL model when exploiting different contextual information, all eight datasets must feature exactly the same $IM_i \rightarrow C_r$ instances, with the only difference being the representation of $IM_i$. 

Since specific contextual information (\eg open issues) cannot be extracted for all instances, once built all datasets we computed their intersection, featuring 85,266 $IM_i \rightarrow C_r$ instances which are present in all datasets.

\subsection{Coding Context}

To extract the \textbf{method calls} context, we run \textsc{java-callgraph}~\cite{javacallgraph} on the corresponding compiled release, thus identifying the $IM_i$'s call graph. As previously explained, methods invoked in the masked part of $IM_i$ have been ignored, since in a real scenario those are the statements that the developer is writing.

Concerning the \textbf{class signatures} context, we relied on the methods previously extracted using \textsc{javalang} \cite{javalang}, appending to $IM_i$ those implemented in its same class as shown in \figref{fig:context}. 

Finally, for the \textbf{most similar method} context, we defined a process to identify, among all methods in the training set, the most similar to $IM_i$ (not considering in $IM_i$ the masked statements). Given the size of the datasets usually adopted to train DL models, we need a scalable and accurate procedure to compute the similarity between a given input method $IM_i$ and all methods in the training set. We start by computing the token-level Jaccard similarity \cite{hancock:dbcb2004} between $IM_i$ and each instance in the training set. Such a metric is very efficient to compute and basically indicates the overlap in code tokens between two methods. Then, we select the top-$k$ methods in the training set which, accordingly to the Jaccard similarity, are the most similar to $IM_i$. Finally, for each of these $k$ methods we compute their CrystalBLEU~\cite{eghbali:icse2022} similarity with $IM_i$, re-ranking them based on this metric. The recently proposed CrystalBLEU has been shown to be the metric better correlating with human assessment when judging the similarity between code snippets. The drawback of this metric is its scalability, which makes it unsuitable to compute the similarity between all instances in the training set and $IM_i$. In summary, we use the Jaccard similarity as a preliminary filter to identify candidate similar methods. Then, we refine such a set using a more reliable metric with the goal of selecting the \emph{most} similar method, being the one augmenting the contextual information. 

In our implementation, we set $k$=20 to achieve a good compromise between scalability and accuracy. While different values may lead to better performance, our goal is not to find the best possible contextual information for code completion, but rather to show that this information can play a substantial role in the model's performance. 


\subsection{Process Context}

The creation of the two process contexts described in \secref{sec:context:process} requires the identification of the $IM_i$ ``most similar issue''. We trained a Transformers and Sequential Denoising Auto-Encoder (TSDAE)~\cite{wang2021tsdae} model for such a task. TSDAE is a denoising auto-encoder based on BERT that can be used to create embeddings. By providing a textual instance to TSDAE, it returns an embedding representing that specific text. We leverage these embeddings to measure the similarity between $IM_i$ and the set of open issues.

To train TSDAE for such a task we built a dataset to make the model learning when an issue is relevant for a given $IM_i$. We used the instances in the ``baseline'' training set as a starting point. As explained, each $IM_i \rightarrow C_r$ instance has been built by looking for the latest commit ($l_c$) that changed the method from which $IM_i$ derived. We indicate the date in which $l_c$ has been performed with $t$. Given an instance, we identify all issues whose status was open at time $t$. Then, we checked if $l_c$ can be ``linked'' to one of the open issues. A link between $l_c$ and an open issue is established if $l_c$'s commit message contains an explicit reference to the issue id (\eg ``\emph{fixed issue \#134}'') or to the issue url (\eg ``\emph{working on issues/134}''). We established such a link for 27,851 instances in our training set. Each of them was used to train TSDAE for the task of identifying the ``open issue'' relevant for a given $IM_i$. Indeed: (i) $l_c$ is the commit that lastly modified the method from which $IM_i$ is derived; (ii) $l_c$ has been performed at time $t$ and can be linked to a specific issue $OI_n$ that was open at that time. As a consequence, we can create one training instance for TSDAE indicating that $OI_n$ is relevant for $IM_i$. Both $IM_i$ and the text composing $OI_n$ are subject to standard pre-processing before they are provided to TSDAE: We exclude Java keywords, remove punctuation, and split camelCase identifiers. 

To choose the best configuration for TSDAE, we performed hyperparameters tuning and experimented the different configurations on a validation set we built starting from the ``baseline'' validation set using the same procedure described for the training set (3,434 instances). We experimented with six different configurations of TSDAE involving 3 different schedulers and 2 different learning rates. Each model has been trained for 4 epochs. The best configuration has been identified as the one having the highest Mean Reciprocal Rank (MRR), indicating the ability of the model to correctly rank in the first positions the issue relevant for $IM_i$. Once identified the best configuration (complete data in our replication package \cite{replication}), we assessed the performance of the trained TSDAE on a test set derived from the ``baseline'' test set (3,256 instances). 

We achieved a MRR of 0.34, which is substantially better as compared to that of a random ranker which, on our test dataset, would obtain a MRR of 0.14.

We create the \textbf{issue title} and \textbf{issue body} context datasets by exploiting the trained TSDAE to identify the most relevant open issue for each $IM_i$. Instances for which no open issues were found at time $t$ are excluded from this dataset and, as a consequence, from all other datasets and our experiment for the reasons previously explained (\ie the need to compare the DL models when trained/tested on exactly the same instances).

\subsection{Developer Context}

The core idea is to provide the model with information characterizing the developer who will receive the completion recommendation. 
In our dataset every $IM_i \rightarrow C_r$ instance has been derived from a $l_c$ commit that impacted the method $IM_i$ by changing the $C_r$ statements. 

Being a commit, $l_c$ has been authored by a developer $D_j$ which, in our study design, is the one who would have received the completion recommendation while working on $IM_i$. Thus, we start by retrieving up to ten past commits performed by $D_j$ before $l_c$ and impacting at least one Java file. We store the diff of these commits as the set of lines of code they added, deleted and modified. Then, we create the \textbf{developer's most similar statements} context by identifying, in this set, the ten statements having the highest similarity with the method $IM_i$ (see \figref{fig:context}). As usual, we do not consider the masked statements (\ie the ones to complete) when computing the similarity. The similarity is based on the percentage of overlapping tokens between each statement and $IM_i$ excluding Java keywords and punctuation. 

Concerning the \textbf{developer's frequent method invocations}, we use srcML~\cite{SrcML} to parse the same set of lines recently added, deleted, or changed by $D_j$ to extract all impacted method calls (both internal or external to the project). Then, we sort them by frequency, keeping up to 100 most frequent calls in the additional context (see \figref{fig:context} for an example).
The choice of keeping only the top-10 most similar statements as compared to the top-100 most frequent calls is due to the fact that entire statements are usually longer than method calls. Thus, given the space available to represent contextual information, we can fit more method calls as compared to entire statements.

%

\section{Study Design} \label{sec:design}

\definecolor{gray50}{gray}{.5}
\definecolor{gray40}{gray}{.6}
\definecolor{gray30}{gray}{.7}
\definecolor{gray20}{gray}{.8}
\definecolor{gray10}{gray}{.9}
\definecolor{gray05}{gray}{.95}

\newlength\Linewidth
\def\findlength{\setlength\Linewidth\linewidth
	\addtolength\Linewidth{-4\fboxrule}
	\addtolength\Linewidth{-3\fboxsep}
}

\newenvironment{rqbox}{\par\begingroup
	\setlength{\fboxsep}{5pt}\findlength
	\setbox0=\vbox\bgroup\noindent
	\hsize=0.95\linewidth
	\begin{minipage}{0.95\linewidth}\normalsize}
	{\end{minipage}\egroup
	\textcolor{gray20}{\fboxsep1.5pt\fbox{\fboxsep5pt\colorbox{gray05}{\normalcolor\box0}}}
	\endgroup\par\noindent
	\normalcolor\ignorespacesafterend}

The \emph{goal} of this study is to assess the impact on the performance of a DL-based code completion technique of additional contextual information provided to it as input. 

We answer the following research question: \textit{To what extent do different types of contextual information impact the performance of  DL-based code completion models?}


We assess ``performance'' by looking at the number of correct predictions generated by the different variations of the DL model (\ie those trained using the different datasets presented in \secref{sec:dataset}, each of which represents a different context). 

In addition to that, we test combinations of the contextual information presented before (\eg \emph{issue title} + \emph{issue body}), for a total of 18 models involved in our study. In the following we detail the DL model we use and how we trained it (\secref{sub:model}), and the process we use to collect and analyze the data output of our study (\secref{sub:data}).

\subsection{DL Model and Training Procedure}
\label{sub:model}

As previously explained, we build on top of the work by Ciniselli \etal \cite{ciniselli:tse2021} that we use as ``baseline''. Thus, we adopt their same DL model, namely the T5 \cite{raffel2019exploring}. T5 has been presented by Raffel \etal~\cite{raffel2019exploring} in five variants characterized by different architectures and, consequently, by a different number of trainable parameters going from 60M for $T5_{small}$ up to 11B for $T5_{11B}$. 
A larger number of parameters implies better performance at the cost of longer training times~\cite{raffel2019exploring}. While Ciniselli \etal \cite{ciniselli:tse2021} opted for the smallest $T5_{small}$, we decided to adopt the $T5_{base}$ (220M), being it more representative of large language models which may be deployed in practice. 


Before being specialized for a task at hand (in our case, code completion), T5 can be pre-trained using a self-supervised task. The goal of the pre-training is to expose the model to the language of interest, making it learning its structure. A typical pre-training objective is the \emph{masked language model}: Assuming the interest in teaching T5 the structure of the Java language, we can provide the model with Java snippets in which 15\% of the tokens composing them have been masked, asking T5 to guess those tokens. Once pre-trained, the model can then be subject to the second training phase, named fine-tuning, in which it is exposed to the specific task of interest. 

Concerning the pre-training, we start from an already pre-trained T5 that has been trained for 1M steps on the C4 dataset~\cite{raffel2019exploring}, featuring 20TB of web-extracted English text. Indeed, previous work showed that starting from a model pre-trained on English is beneficial when dealing with code as compared to the randomly initialized weights of a non pre-trained model \cite{Tufano:arxiv2020}. Starting from this checkpoint, we additionally pre-train the model for 500k steps using the previously described \emph{masked language model} objective on a Java pre-training dataset we built. The dataset features 12,671,475 Java methods that have been extracted from GitHub projects also in this case identified using the search platform by Dabic \etal~\cite{Dabic:msr2021data}. Also in this case we targeted non-forked projects with a long change history ($>$500 commits), at least a small development team ($>$10 contributors), and at least 10 stars. Java methods containing non-ASCII characters, being longer than 512 tokens, or being already present in the fine-tuning datasets have been excluded.


	
\begin{table}[ht]
	\centering
	\caption{Fine-tuning Datasets.}
	\scriptsize
	\label{tab:datasets}
	\begin{adjustbox}{width=\columnwidth,center}
	\begin{tabular}{lrrr}
	\toprule
	\multirow{2}{*}{{\bf Context}}  & \multicolumn{3}{c}{\bf Instances Length}\\ \cline{2-4}
	& {\bf Mean} & {\bf Median} & {\bf St. Dev.}\\\midrule
Baseline  & 243 & 205 & 166\\
Method Calls (MC)  & 380 & 326 & 253\\
Class Signatures (CS)  & 733 & 481 & 1,128\\
Most Similar Method (MSM) & 447 & 384 & 296\\
Issue Title (IT)  & 260 & 222 & 167\\
Issue Body (IB) & 550 & 361 & 1,429\\
Frequent Invocations (FI)  & 467 & 418 & 252\\
Most Similar Statements (MSS) & 517 & 445 & 2,355\\\midrule

MSM + CS  & 941 & 693 & 1,169\\
MC + CS  & 871 & 621 & 1,153\\
MSM + MC  & 584 & 507 & 374\\
MSM + MC + CS  & 1,079 & 829 & 1,200\\
IT + IB  & 567 & 378 & 1,430\\
FI + MSS   & 741 & 647 & 2,365\\\midrule
Best Code + Best Process  & 601 & 525 & 376\\
Best Code + Best Developer  & 808 & 735 & 423\\
Best Developer + Best Process  & 484 & 435 & 254\\
Best Code+ Best Developer + Best Process  & 825 & 753 & 425\\

\bottomrule
\end{tabular}
\end{adjustbox}
\end{table}

The fine-tuning datasets are the ones described in \secref{sec:dataset} plus their combinations as reported in \tabref{tab:datasets}. All datasets are composed by instances in the form $IM_i \rightarrow C_r$, in which $IM_i$ is the method to complete possibly augmented with contextual information and $C_r$ the expected completion. \tabref{tab:datasets} also reports statistics about the length of the instances (in terms of number of tokens) in each dataset. 

All fine-tuning datasets feature 85,266 instances, split into 80\% training (68,215), 10\% evaluation (8,526), and 10\% test set (8,525). While the datasets have been randomly split, all instances referring to the same method belong to the same set, to avoid biasing our results. Indeed, it is worth remembering that a method may generate multiple instances in our dataset, since we could mask different parts of it.

As shown in \tabref{tab:datasets}, we experiment with: (i) the baseline model; (ii) the seven types of context introduced in \secref{sec:dataset}; (iii) all combinations of contexts within the same family (\eg combinations of the three \emph{coding contexts}; and (iv) combinations of contexts across different families (\eg combining a \emph{coding context} with a \emph{process context}). For these cross-families combinations, we reduce the number of experiments to run by only considering the best combination within each family. This means that we combine the best \emph{coding context} (which could be a single context or a combination of multiple \emph{coding contexts}) with the best \emph{process context}; then, we combine the best \emph{coding context} with the best \emph{developer context}; \etc We assess what the best context is within each family by looking at the number of correct predictions (\ie recommended code is identical to the expected one) generated by the models.

\textbf{Training procedure.} We pre-trained and fine-tuned T5 using a Google Colab's 2x2 TPU topology (8 cores)~\cite{colab}. We also trained a 32k word-pieces SentencePiece tokenizer~\cite{kudo2018sentencepiece} used by the model to represent the input/output. The tokenizer has been trained on 1M Java methods randomly extracted from the pre-training dataset and 712,634 English sentences from C4~\cite{raffel2019exploring}. The maximum number of tokens for the input has been set to 1,024 and the batch size to 32.

We performed hyperparameters tuning assessing the performance of the four different T5 configurations experimented by Ciniselli \etal \cite{ciniselli:tse2021}. These configurations differ for the way they handle the learning rate. We fine-tuned each configuration for 30k steps and assessed its performance on the evaluation set in terms of its ability to generate correct predictions. To reduce the cost of such a procedure, we found the best configuration only on the ``baseline'' dataset (\ie no additional contextual information provided), and used it in all experiments. The best configuration found was the one using a constant learning rate equal to 0.001. Such a configuration has been used to fine-tune the 18 different models (\ie different contexts and combinations of contexts). Each model has been fine-tuned for 160k steps, corresponding to $\sim$75 epochs on the 68,215 instances of the training dataset. To avoid overfitting, we saved a checkpoint for each model every 5k training steps. Then, we evaluated the performance of the different checkpoints on the evaluation set, picking the best one as the final model to use in our experiments on the test set.

\subsection{Data Collection and Analysis}
\label{sub:data}

We run the $18$ trained models on the test set assessing their performance in terms of correct predictions. We considered a prediction to be correct if it matches the expected code, except for differences in spaces (\eg two \emph{vs} one space between two tokens). 

To evaluate whether the difference in correct predictions generated by two models is statistically significant, we use the McNemar's test \cite{mcnemar}, useful to compare dichotomous results of two different treatments, together with the Odds Ratio (OR) effect size. We account for multiple comparisons by adjusting $p$-values using the Holm's correction \cite{Holm1979a}.

We also assess the complementarity between the baseline and each contextual model by computing the percentage of correct predictions generated by (i) both models (\ie for a given code completion instance, both models correctly recommend the completion); (ii) the baseline only; (iii)  the contextual model only. This analysis allows to understand the potential of combining multiple models.
\section{Results} \label{sec:results}
\tabref{tab:performance} reports the percentage of correct predictions generated by T5 when trained using the \emph{baseline} representation, the different types of contexts, and their combinations (within- and cross-family). The baseline achieves 30.58\% of correct predictions which is inline with what observed by Ciniselli \etal \cite{ciniselli:tse2021} when investigating the ability of T5 to generate entire statements (in our case, up to two statements). When providing the model with additional contextual information of a specific type (Context Type = ``Single'' in \tabref{tab:performance}), we observe an increase in performance ranging from a relative +0.6\% (\emph{issue body}) to a +7\% (\emph{most similar method}). Also providing the model with the \emph{method calls} context (\ie the methods invoking and invoked by the method to complete) provides a substantial boost in correct predictions (+6\%).



\begin{table}[ht]
	\centering
	\caption{Percentage of correct predictions.\vspace{-0.2cm}}
	\scriptsize
	\label{tab:performance}
	\begin{adjustbox}{width=\columnwidth,center}
	\begin{tabular}{llr}
	\toprule
	{\bf Context} &  \multirow{2}{*}{\bf Context}  & {\bf \%Correct}\\ 
	{\bf Type} &  & {\bf Prediction}\\\midrule
	     None                 & Baseline                             & 30.58\% \\\midrule
	                              & Method Calls (MC)                      & 32.46\% \\
	                     & Class Signatures (CS)                        & 31.33\% \\
	                      &  Most Similar Method (MSM)        & 32.68\% \\
	     Single                  & Issue Title (IT)                     & 31.32\% \\
	          		& Issue Body (IB)                        & 30.77\% \\
	          & Frequent Invocations (FI)                        & 31.80\% \\        
	             	         &  Most Similar Statements (MSS)                     & 31.00\% \\\midrule 
	                     
\multirow{6}{*}{Within-family} 	         & MSM + CS                         & 32.72\% \\
	                  & MC + CS                     & 33.03\% \\
	                      &  MSM + MC & 33.35\% \\
	                      &  MSM + MC + CS        & 33.88\% \\
		                      &  IT + IBy        & 31.20\% \\	                      
	                      &   FI + MSS        & 31.17\% \\\midrule	 
	  \multirow{4}{*}{Cross-family}          & Best Code + Best Process                      & 33.75\% \\
	            & Best Code + Best Developer& 33.58\% \\
	                      &  Best Developer + Best Process        & 31.50\% \\
	                      &  Best Code+ Best Developer + Best Process        & 33.54\% \\
\bottomrule
\end{tabular}
\end{adjustbox}
\end{table}

\begin{figure}[ht]
   \centering
    \includegraphics[width=\linewidth]{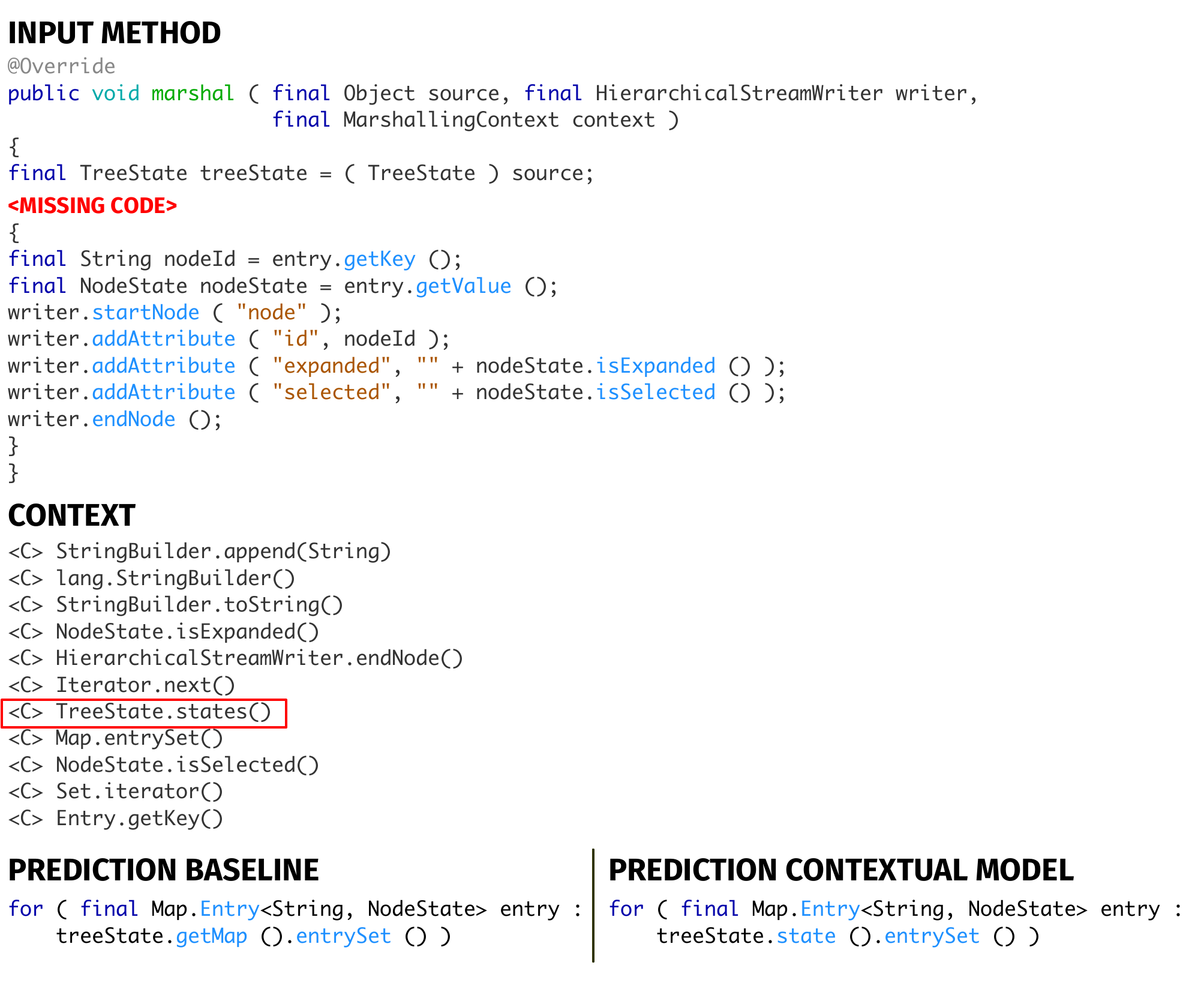}
    \caption{\emph{Method calls} context helping the prediction.}
    \label{fig:methodCalls}
\end{figure}


When statistically comparing the performance of the \emph{baseline} with the models exploiting individual types of contextual information, we found significant differences ($p$-value $<$ 0.05) in all cases but for \emph{issue body} and \emph{most similar statements}. The OR for the significant differences ranges between 1.23 (\emph{class signatures}) and 1.64 (\emph{method calls}). 

An OR=1.64 indicates 64\% higher odds of obtaining a correct prediction using the contextual model as compared to the \emph{baseline}. The complete statistical analysis is available in our replication package \cite{replication}.

\figref{fig:methodCalls} provides an example in which the \emph{baseline} model was not able to generate the expected completion, wrongly recommending \texttt{treeState.\-getMap} in the \texttt{for} loop. When providing the model with \emph{method calls} context, the model was able to exploit such information to identify \texttt{treeState.\-state} as the correct method call to feature in the \texttt{for}. 


Worth mentioning is also the +4\% ensured by the \emph{frequent invocations} context (OR=1.44), featuring the method calls frequently used by the developer receiving the recommendation. 

Among the \emph{process contexts}, feeding the model with the title of the most relevant open issue seems to help (+2.5\%), while this is not the case when the issue body is provided.

By combining the contexts belonging to the same family (Context Type = ``Within-family'' in \tabref{tab:performance}), the improvement in performance can be pushed further when it comes to the \emph{coding contexts}. The three \emph{coding contexts} together result in a relative +11\% in correct predictions as compared to the baseline (33.88\% \emph{vs} 30.58\%) --- $p$-value $<$0.0001, OR=1.9. Such an improvement is not obtained, instead, for the other two families of contexts (\ie \emph{process} and \emph{developer}), for which the performance of the combinations are inline of slightly worse than the single context types taken in isolation. 

The bottom part of \tabref{tab:performance} reports the results achieved by combining the contexts being the best performing of each of the three families. This includes the \emph{issue title} as representative of the \emph{process context}, and the \emph{frequent method invocations} for the \emph{developer context}. Concerning the \emph{coding context} a longer discussion is needed: The best performing model is the one exploiting a combination of all information (\ie \emph{most similar method + method calls + class signatures}). However, such a context tends to saturate the 1,024 tokens available for the model's input. Thus, combining it with even additional contextual information would not make sense, since the input will be cut in most of cases. 

For this reason, we selected the second best-performing model among the \emph{coding contexts}, namely \emph{most similar method + method calls}. The latter, while ensuring performance similar to the best one (33.35\% \emph{vs} 33.88\% of correct predictions) requires, on average, half of the tokens for its representation. 

As it can be seen from \tabref{tab:performance}, while improvements can be obtained in terms of correct predictions as compared to the \emph{baseline} ($p$-value $<$ 0.0001 in all comparisons, with ORs ranging between 1.31 and 1.9), none of the experimented cross-family combinations outperforms the best within-family combination featuring all coding contexts. Such a result may be due to the fact that we did not manage to experiment with cross-family combinations involving the best within-family context (due to limitations in the input size).\smallskip

\textbf{Take-away.} \emph{Additional contextual information can have a substantial impact on the model's performance. Our experiments showed relative improvements up to +11\% in terms of correct predictions.}


\subsubsection*{\textbf{Complementarity Analysis and Confidence of the Predictions}}
\tabref{tab:complementarity} reports the results of the complementarity analysis concerning the correct predictions generated by the \emph{baseline} and by the models using different combinations of contextual information. Given the \emph{baseline} and a specific context $C_i$, this means computing the union of the correct predictions generated by both approaches, and then counting those (i) generated by both models (\ie both models generated a correct prediction for a given instance), (ii) generated by the \emph{baseline} only, and (iii) generated by the model exploiting $C_i$ only. For example, in the case of the \emph{method calls} context, 78.12\% of correct predictions are shared with the baseline, 8.29\% are only generated by the \emph{baseline}, and 13.59\% are only generated by the model exploiting the \emph{method calls} context. 



\begin{table}[ht]
	\centering
	\caption{Complementarity analysis: between the correct predictions (CP) generated by the \emph{baseline} and by the models exploiting different contextual information.}
	\scriptsize
	\label{tab:complementarity}
	\begin{adjustbox}{width=\columnwidth,center}

	\begin{tabular}{lrrr}
	\toprule
	\multirow{2}{*}{\bf Context} &  {\bf \%CP}  &  {\bf \%CP Only} &  {\bf \%CP Only}\\
	 & {\bf  Shared}  & {\bf Baseline} & {\bf Context}\\\midrule
	                     Method Calls (MC)                     & 78.12\% & 8.29\% & 13.59\% \\
	                     Class Signatures (CS)                        & 79.03\% & 9.4\% &11.57\% \\
	                     Most Similar Method (MSM)        & 73.8\% & 10.22\% & 15.98\% \\
	                      Issue Title (IT)                   & 82.85\% & 7.48\% & 9.67\% \\
	                      Issue Body (IB)                       & 82.74\% & 8.35\% & 8.91\% \\	                     
	                      Frequent Invocations (FI)                        & 80.45\% & 8.01\% & 11.54\%\\
	                      Most Similar Statements (MSS)                     & 79.79\% & 9.49\% & 10.72\%\\\midrule

	                     MSM + CS  & 72.62\% & 10.78\% & 16.6\% \\
	                     MC + CS       & 75.73\% & 8.75\% &15.52\% \\
	                     MSM + MC & 72.52\% & 10\% & 17.48\% \\	                    
	                     MSM + MC + CS        & 71.72\% & 9.75\% & 18.53\% \\
	                      IT + IB        & 81.68\% & 8.24\% & 10.08\% \\	                      
	                      FI + MSS        & 78.01\% & 10.15\% & 11.84\% \\\midrule	 
	                      Best Code + Best Process                      & 72.72\% & 9.39\% & 17.89\% \\
	                      Best Code + Best Developer		& 73.27\% & 9.31\% & 17.42\\
	                      Best Developer + Best Process       & 80.37\% & 8.48\% & 11.15\% \\
	                      Best Code+ Best Developer + Best Process        & 71.46\% & 10.32\% & 18.22\% \\

\bottomrule
\end{tabular}
\end{adjustbox}
\end{table}

\eject

The results in \tabref{tab:complementarity} provide one important message: There is a good complementarity between the \emph{baseline} and the models exploiting additional contextual information. For example, while the best-performing model (\ie \emph{most similar method + method calls + class signatures}) generates 18.53\% of correct predictions which are missed by the \emph{baseline}, the latter is still able to generate 9.75\% of correct predictions which are missed by the contextual model. 

Such a result points to the possibility of combining multiple models exploiting different input representations to boost performance. One possibility is to trigger, for a given code completion scenario, the model having the highest confidence in its prediction. Indeed, as most of DL models, T5 provides a \emph{score} for each generated prediction. The score is a value lower than 0 representing the log-likelihood of the prediction. For example, having a log-likelihood of -1 means that the prediction has a likelihood of 0.37 ($ln(x)=-1 \Longrightarrow x=0.37$). The likelihood can be interpreted as the confidence of the model about the correctness of the prediction on a scale from 0.00 to 1.00 (the higher the better).

\begin{figure*}[t]
   \centering
   \vspace{-0.3cm}
    \includegraphics[width=0.78\textwidth]{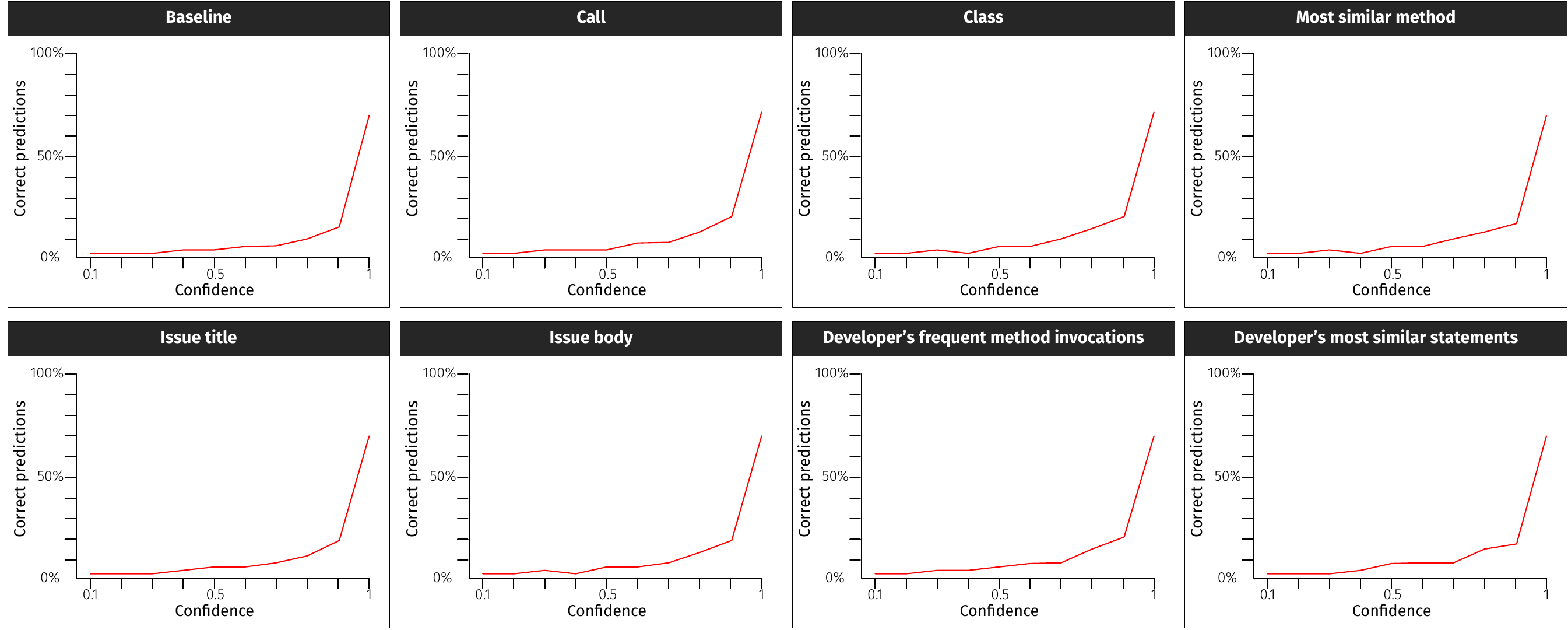}
    \caption{Percentage of correct predictions when varying the confidence of the model.}
    \label{fig:confidence}
    \vspace{-0.3cm}
\end{figure*}

\figref{fig:confidence} shows the relationship between the percentage of correct predictions ($Y$-axis), and the T5 confidence ($X$-axis). We grouped the predictions into different buckets based on their confidence (\ie from 0.00 to 0.10, from 0.11 to 0.20, $\dots$, from 0.91 to 1.00). The results are shown for the \emph{baseline} and for the models exploiting each contextual information in isolation. There is a clear trend indicating that the higher the confidence of the prediction, the higher the likelihood of obtaining a correct prediction. When the confidence is greater than 0.90, the models are usually able to recommend the correct completion in more than 70\% of cases. 

Given the complementarity observed for the different models and the reliability of the confidence as a ``proxy'' for the prediction quality, we experimented with a \emph{confidence-based} model which, given a code completion scenario from our test set, recommends as completion the output of the model having the highest confidence among all those we experimented with (\ie the \emph{baseline} and the ones exploiting different combinations of contextual information). \tabref{tab:score} shows in the top part a performance comparison between the \emph{baseline} and the \emph{confidence-based} model in terms of correct predictions. As it can be seen, the \emph{confidence-based} model is by far the best we experimented with, increasing the percentage of correct predictions generated by the \emph{baseline} by a relative +22.4\% (from 30.58\% to 37.43\%). This results in a statistically significant difference ($p$-value $<$ 0.0001) with an OR=6.56. 



\begin{table}[ht]
	\centering
	\caption{Baseline \emph{vs} confidence-based model.\vspace{-0.2cm}}
	\scriptsize
	\label{tab:score}
	\begin{tabular}{lrr}
	\toprule
	{\bf Measure} & {\bf Baseline} & {\bf Conf. model}\\\midrule
	\multicolumn{3}{c}{PERFORMANCE COMPARISON}\\\midrule
	Correct Predictions (\#) & 2,607 & 3,191\\
	Correct Predictions (\%) & 30.58 & 37.43\\\midrule
	\multicolumn{3}{c}{COMPLEMENTARITY ANALYSIS}\\\midrule
	Exact Match Prediction Shared& \multicolumn{2}{c}{2,502/3,296 (75.91\%)}\\
	Exact Match Predictions only Baseline& \multicolumn{2}{c}{105/3,296 (3.19\%)}\\
	Exact Match Predictions only Score model& \multicolumn{2}{c}{689/3,296 (20.90\%)}\\

\bottomrule
\end{tabular} 
\end{table}

Finally, the bottom part of \tabref{tab:score} shows the complementarity analysis between the \emph{baseline} and the \emph{confidence-based} model. As it can be seen, there is a 20.9\% of correct predictions only generated by the \emph{confidence-based} model. Only a 3.19\% of correct predictions is instead generated only by the \emph{baseline} model. 

\begin{figure}[ht]
   \centering
    \includegraphics[width=\linewidth]{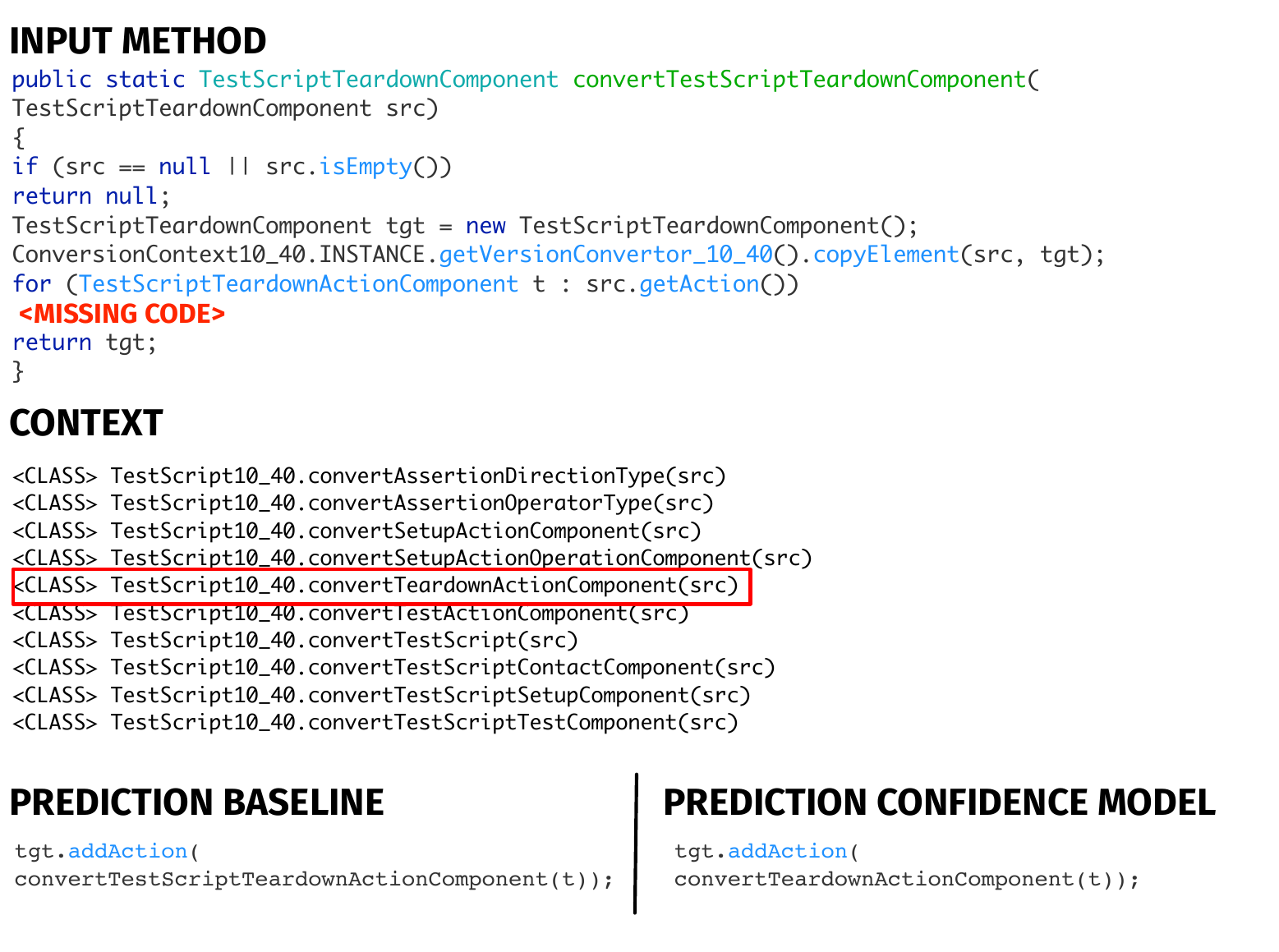}
    \caption{\emph{Class signatures} context helping the prediction of the \emph{confidence-based} model.\vspace{-0.4cm}}
    \label{fig:classSignature}
\end{figure}

We conclude this section with a practical example that shows how the \emph{confidence-based} model can take advantage of the context most suitable in each situation. \figref{fig:classSignature} shows a Java method performing a conversion of a specific component. The \emph{baseline} model is able to correctly understand the meaning of the function and it is aware of the need to call the conversion method on the object \emph{t} but lacks context that could suggest the method's name, thus making an incorrect prediction. Conversely, the \emph{confidence-based} model understands that in this situation the \emph{class signatures} context can be exploited, finding out the possibility to access the method \emph{convertTeardownActionComponent}, resulting in a correct prediction.

\textbf{Take-away.} \emph{Combining models using different contextual information by exploiting the confidence of their prediction can provide substantial benefits in terms of correct predictions for code completion.}

\section{Validity Discussion} \label{sec:threats}

\textbf{Construct validity.} The usage of the correct predictions metric provides a limited view about the performance of the experimented models. For example, a model may generate a code completion prediction different but behaviorally equivalent to the expected one. Our experimental design simply considers such a prediction as wrong. However, given the goal of our study, we preferred to use a single and easy to interpret metric. For example, when comparing techniques it might be difficult to interpret what a +3\% in terms of average CrystalBLEU~\cite{eghbali:icse2022} means in practice. 

\eject

\textbf{Internal validity.} An important factor influencing DL performance is the calibration of hyperparameters which, due to feasibility reasons, we limited to the \emph{baseline}, assuming the others would benefit from the same configuration. However, a hyperparameters tuning also extended to the contextual models may only improve the performance of the latter, thus further reinforcing our finding: Additional contextual information can have a substantial impact on the model's performance.

Our experiment only focuses on ``expanding'' the contextual information provided to the model as compared to the \emph{baseline}. One may wonder if good results can be obtained by, instead, shrinking the input representation. We experimented with this scenario, by creating two representations of the method to complete in which, given $S$ the set of masked statements, we only provide the model with up to six or four statements surrounding it (rather than the whole method containing $S$). To better understand, the representation using up to six surrounding statements inputs to the model $S$ (the masked part to generate) with up to three statements above and below it. We say ``up to'', since not in all cases there will be at least three statements above/below $S$. We found that shrinking the contextual information provided to the model results in a strong drop of performance, with a relative loss in terms of perfect predictions of -19.48\% and -22.24\% when providing only the six and the four surrounding statements, respectively. Complete results are available in our replication package \cite{replication}.

\textbf{Conclusion validity.} As explained in \secref{sub:data} we used appropriate statistical procedures, also adopting \emph{p}-value adjustment when multiple tests were used within the same analysis. The differences in performance among the different models might look minor at a first sight. For example, the \emph{confidence-based} model achieves 37.43\% of correct predictions versus the 30.58\% of the \emph{baseline}, resulting in a +22.4\% relative improvement but \emph{only} in a +6.85\% absolute improvement. 

Even the latter actually is a major improvement as compared, for example, to previous work in the literature proposing novel code completion techniques (\eg +0.8\% in accuracy, Table 3 in \cite{izadi:icse2022}).

\textbf{External validity.} We used T5 as representative of DL-based code completion techniques \cite{ciniselli:tse2021}. Other DL models may lead to different results. Also,  we targeted Java and statement-level code completion (\ie completing up to two complete statements). Our findings may not generalize to other settings.
\section{Related Work} \label{sec:related}

While several code completion techniques have been proposed in the literature \cite{bruch:fse2009,proksch:tosem2015,robbes:ase2010,hindle:icse2012,asaduzzaman:icsme2014,tu:fse2014,hellendoorn:fse2017}, given the goal of our study, we only focus on those exploiting DL. 

Karampatsis \etal~\cite{karampatsis:arxiv2019} proposed the use of \textit{Byte Pair Encoding} (BPE) \cite{gage:cusers1994} when applying neural networks to the task of code completion. This allows to overcome the out of vocabulary problem (\ie the impossibility of neural network model to keep memory of the huge number of words in a corpus). They showed that, using BPE, DL models are the best choice for tackling code completion.

Alon \etal~\cite{alon:icml2020} proposed Structural Language Model (SLM), a language agnostic approach leveraging the AST to represent the statement with the missing tokens to complete. Their architecture, that combines LSTMs and Transformers, was able to correctly recommend completion in 18.04\% of cases with a single attempt.

Differently from \cite{alon:icml2020}, Svyatkovskiy \etal~\cite{svyatkovskiy:fse2020} did not exploit any structural representation for the code, treating it like a stream of tokens. Their model leveraged the Transformers architecture and BPE \cite{gage:cusers1994} to recommend even an entire statement, achieving a perplexity of 1.82 for a Python corpus.

A Transformer-based architecture was also proposed by Ciniselli \etal~\cite{ciniselli:tse2021}. The authors compared two different Transformer-based models, the T5, and the RoBERTa model, with an $n$-gram model when completing blocks of code with up to two entire statements. Their best model, the T5 model, was able to correctly predict 29\% of the blocks. This model represents the \emph{baseline} in our experiments.

Feng \etal \cite{feng:emnlp2020} proposed CodeBERT, a bimodal Transformer trained on code and English text. Having been trained using a ``masking''  pre-training objective, CodeBERT is suitable for code completion, despite the authors focus on the problems of code search and code documentation.

Wang \etal \cite{wang:emnlp2021} presented CodeT5, in which the T5 has been pre-trained with a novel identifier-aware task. The semantic information allowed CodeT5 to achieve state-of-the-art performance on the CodeXGLUE benchmark.

Izadi \etal \cite{izadi:icse2022} presented CodeFill, a Transformer-based approach trained for single- and multi-token code completion by predicting both the type and value of the masked tokens. CodeFill outperforms previous techniques.

Chen \etal \cite{chen:arxiv2021} introduced GitHub Copilot, a DL model trained on 159Gb of data from GitHub. The model achieved unprecedented capabilities, being able to even predict the entire method given the description of the task to perform. 

Related to our work are also studies proposing the use of contextual information for improving recommender systems for developers. Gail Murphy suggested that contextual information ``\emph{could enable software tooling to take a substantial step forward}'' \cite{murphy:icse2019}. Tian and Treude \cite{tian:icsme2022} presented a preliminary study in which they evaluated how additional contextual information provided as input to a DL model may improve its performance for clone detection and code classification. They observed improvements of up to 8\%. Their context is similar to our \emph{method calls}. Our work exploits a larger variety of contexts, addresses a different task, and shows how major improvements can be obtained by combining contextual models in a \emph{confidence-based} approach. 

Similarly, the following discussed works exploit contextual information in the context of developers' recommenders having, however, a focus on other tasks. 

Zhao \etal \cite{zhao:tse2023} introduced APIMatchmaker, a tool able to leverage a multi-dimensional, context-aware, and collaborative filtering approach to recommend API usages by learning from real-world Android apps. They evaluated their approach on 12,000 apps showing state-of-the-art performance, being able to correctly recommend APIs in over 50\% of the cases with just one attempt.
Abid \etal \cite{abid:fse2022} leveraged contextual data from the developer's active project to recommend method bodies extracted from similar projects. They showed the effectiveness of the context involving frequently occurring API usages, achieving a Precision@5 of 94\%.
Wen \etal \cite{Wen:icse2021} exploited association rules to extract ``implementation pattern'' (\ie groups of method usually implemented in the same task). They processed the current code the developer is writing in order to identify an existing implementation pattern and hence recommending the missing method belonging to the same pattern.
%
Asaduzzaman \etal \cite{asaduzzaman:jsep2016} proposed Context-sensitive Code Completion (CSCC), an approach that leverage method calls, java keywords, and class/interface names within the previous 4 lines of code for recommending APIs. Their approach outperformed state-of-the-art tools, achieving recall and precision of 84\% and 86\% respectively, being also able to recommend the suggestion in less than 2ms.
D'Souza \etal \cite{souza:scam2016} presented PyReco, a code completion system for Python that exploit a nearest neighbor classifier to sort the suggested APIs based on the relevance rather than the conventional alphabetic order. Thanks to the rich contextual information collected, like libraries imported and the API methods or attributes, they were able to achieve a Recall of 84\%.
%
%

\section{Conclusion and Future Work} \label{sec:conclusion}
We investigated how augmenting the contextual information provided to a DL model can benefit its performance in the context of code completion. We experimented with three families of contexts, namely \emph{coding context}, \emph{process context}, and \emph{developer context}, showing that they can boost the correct predictions of the \emph{baseline} up to a relative +11\%. Also, the models exploiting different contextual information exhibit a good complementarity. For this reason, we combined them by exploiting the confidence of their predictions (\ie for a given code completion scenario the recommendation is triggered by the model having the highest confidence). This allowed to achieve a relative improvement of +22\% over the \emph{baseline}.

Future work will mostly point to the generalizability of our findings to different languages and code-related tasks.

\section*{Acknowledgment}
This project has received funding from the European Research Council (ERC) under the European Union's Horizon 2020 research and innovation programme (grant agreement No. 851720).

\bibliographystyle{IEEEtranS}
\bibliography{main}

\end{document}